\documentclass[aps,prl,twocolumn,superscriptaddress]{revtex4-2}

\usepackage[british]{babel}
\usepackage[babel=true]{csquotes}
\usepackage{graphicx}
\usepackage{float}

\usepackage[colorlinks=true,linkcolor=blue,citecolor=blue,urlcolor=blue]{hyperref}

\usepackage{amsmath}
\usepackage{amssymb}
\usepackage{mathtools}
\usepackage{physics}
\usepackage{siunitx}
\usepackage{booktabs}

\usepackage[nopostdot,style=super,nonumberlist,toc,nogroupskip]{glossaries}
\glsdisablehyper

\newacronym{soi}{SOI}{silicon-on-insulator}
\newacronym{hom}{HOM}{Hong--Ou--Mandel}
\newacronym{snspd}{SNSPD}{superconducting nanowire single-photon detector}
\newacronym{spdc}{SPDC}{spontaneous parametric down-conversion}
\newacronym{ppktp}{ppKTP}{periodically poled potassium titanyl phosphate}
\newacronym{hwp}{HWP}{half-wave plate}
\newacronym{qwp}{QWP}{quarter-wave plate}
\newacronym{pbs}{PBS}{polarising beam splitter}
\newacronym{mmi}{MMI}{multi-mode interferometer}
\newacronym{gc}{GC}{grating coupler}
\newacronym{mzi}{MZI}{Mach-Zehnder interferometer}
\newacronym{2dc}{2DGC}{two-dimensional grating coupler}
\newacronym{fst}{FST}{full quantum state tomography}
\newacronym{mle}{MLE}{maximum likelihood estimation}
\newacronym{box}{BOX}{buried oxide layer}
\newacronym{tox}{TOX}{top oxide layer}

\newcommand{\affiliationFMQ}{\affiliation{Institute for Functional Matter and Quantum Technologies, University of Stuttgart, 70569 Stuttgart, Germany}}
\newcommand{\affiliationIQST}{\affiliation{Center for Integrated Quantum Science and Technology (IQST)}}

\begin{document}

\renewcommand{\figurename}{Fig.}

\title{Integrated high-fidelity preparation and analysis of photonic two-qubit states for quantum network nodes}

\author{Jonas C. J. Zatsch}
\author{Tim Engling}
\author{Jeldrik Huster}
\author{Louis L. Hohmann}
\author{Shreya Kumar}
\author{Stefanie Barz}
\email[Corresponding author. Email: ]{barz@fmq.uni-stuttgart.de}
\affiliationFMQ{}
\affiliationIQST{}

\date{\today}

\begin{abstract}
	The realisation of quantum networks requires local quantum information processing at the network nodes and highly efficient transmission of quantum information across the network.
    Integrated photonics, based on \acrlong{soi}, is a promising platform for quantum network nodes, as it supports low-loss propagation of telecom wavelength photons, making it compatible with existing optical fibre networks.
	Here, we present a \acrlong{soi} integrated photonic chip, capable of bidirectional operation, enabling the preparation of arbitrary single- and two-qubit states, and performing \acrlong{fst} on up to two qubits.
    Using our chip, we obtain preparation fidelities above $\SI{97}{\%}$ for on-chip prepared Bell states coupled into optical fibres.
	Furthermore, we demonstrate that we can distribute entanglement between network nodes by preparing a two-qubit cluster state on the first node and performing \acrlong{fst} on the second node, achieving a fidelity of $\SI{90.0 +- 1.6}{\%}$.
	This result proves that our approach allows the distribution of entanglement from one chip to another.
	The potential of bidirectional operation makes our circuit a versatile node in telecom quantum networks, both functioning as a sender and receiver unit, a key element for the deployment of fully photonic multi-purpose quantum networks.
\end{abstract}

\maketitle

\section{Introduction}
Integrated quantum photonics, based on the \gls{soi} material stack, is an ideal platform for networked quantum applications.
It offers a robust and scalable implementation of qubit preparation and measurement, due to its small footprint components and low-loss waveguides~\cite{Koen_2025}.
Additionally, \gls{soi} integrated circuits are CMOS compatible, allowing for cost-efficient and high-quality fabrication of quantum processors~\cite{Koen_2025}.
Since it is possible to operate \gls{soi} integrated circuits in the optical C-band, existing low-loss telecom fibre networks can be utilised as quantum channels in a quantum network~\cite{Valivarthi_2016,Bunandar_2018,Du_2024}.

Quantum network nodes capable of preparing, manipulating and analysing photonic qubits are highly desirable.
Implementations of quantum key distribution protocols using integrated photonics have been shown~\cite{Sibson_2017,Zhang_2019,Semenenko_2020,Avesani_2021,Li_2023,Heo_2025,Lin_2025}.
Furthermore, it has already been demonstrated that multi-qubit states can be prepared on a photonic chip with a high fidelity~\cite{Silverstone_2015,Qiang_2018,Adcock_2019}.
Additionally, the coupling of on-chip single-qubit and entangled states to an off-chip quantum channel, an optical fibre, has been realised~\cite{Olislager_2013,Wang_2016,Llewellyn_2020,Liu_2022,Miloshevsky_2024,Koen_2025,Jiang_2025,Hua_2025,Qin_2025}. 
However, an open challenge is the preparation and analysis of path-encoded photonic quantum states of more than one qubit which are coupled to a quantum channel using a single device, while still achieving high fidelities. 
Even more, it remains to be shown, that a maximally entangled path-encoded state can be distributed from one chip to another.

Here, we present an integrated \gls{soi} photonic circuit capable of preparing arbitrary single- and two-qubit product states, as well as maximally entangled two-qubit states encoded into two photons.
Furthermore, these two-qubit states are coupled from the chip to single-mode fibres.
Here, the state encoding is switched from path-encoded qubits to polarisation-encoded qubits, while maintaining the prepared state with a high fidelity of above $\SI{97}{\%}$.
We demonstrate its functionality by preparing three different unentangled two-qubit states and all four Bell states on-chip, coupling them to optical fibres and analysing the states using off-chip \gls{fst} units.
Furthermore, we show that the chip can also function as a receiving node in a quantum network by using it as a two-qubit \gls{fst} unit and performing \gls{fst} on an off-chip prepared Bell state.
For all cases, we measure fidelities above $\SI{97}{\%}$.
Finally, we realise a proof-of-concept network consisting of two nodes by connecting two copies of our chip to each other.
We prepare a maximally entangled state on the sending node, distribute it to the receiving node, where we perform \gls{fst}.
We measure a fidelity of $\SI{90.0 +- 1.6}{\%}$, proving that our chip is capable of sharing entanglement inside the network.
Our photonic chip design enables the realisation of versatile quantum network nodes, capable of acting as a sending unit of entangled quantum states, and also as a receiving unit.

\section{Experimental setup}
\begin{figure*}
	\includegraphics[scale=0.6]{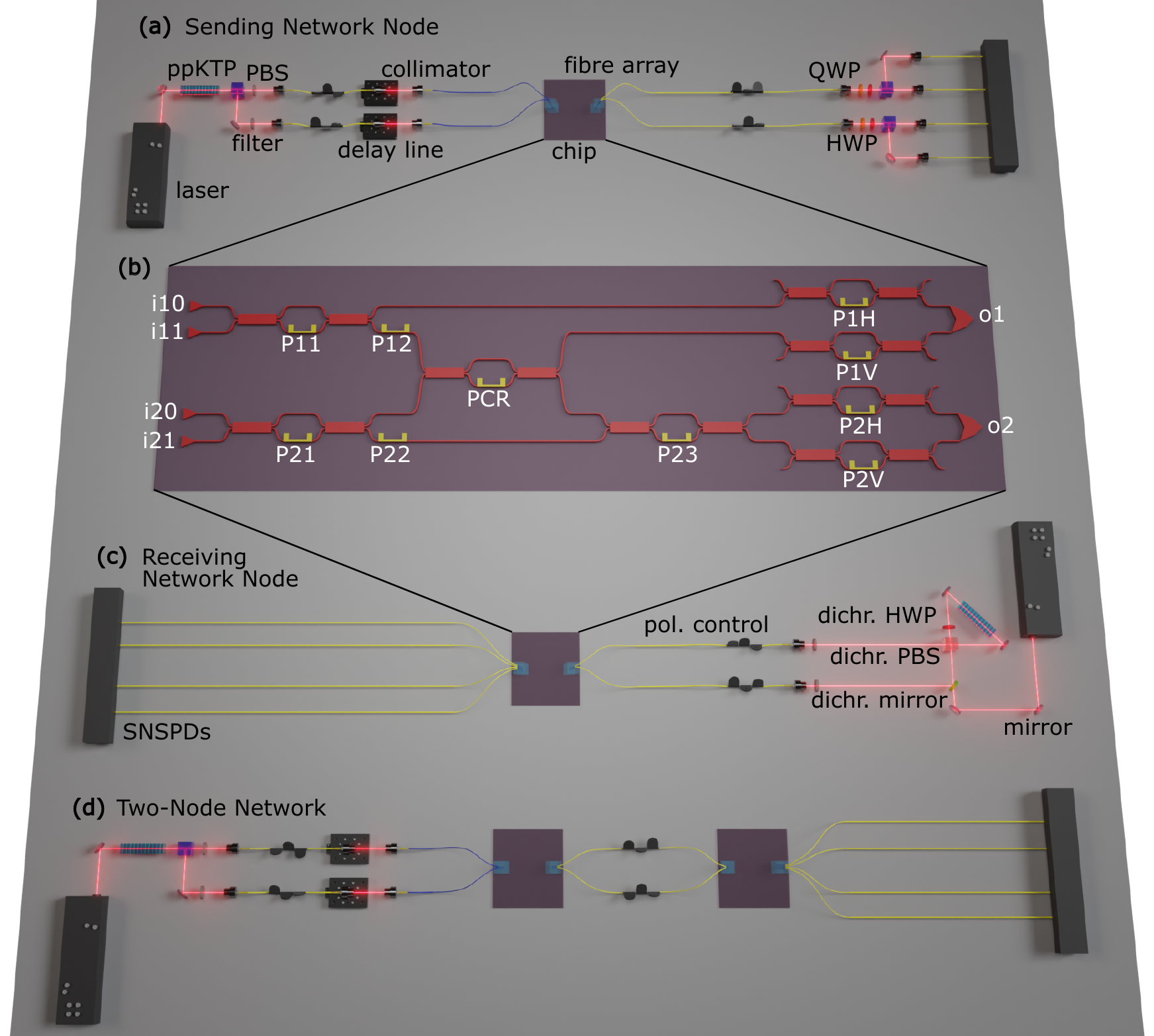}
	\caption{
        \label{fig:setup/fullsetup}
		\textbf{(a)} Schematic of the setup for on-chip state preparation.
		Single photons are generated by an \acrfull{spdc} source consisting of a pulsed $\SI{775}{nm}$ pump laser and a \acrlong{ppktp} crystal.
		A \gls{pbs} is used to split the generated photon pairs and the photons are coupled into polarisation-maintaining fibres.
		The photons are coupled to the chip with the help of a fibre arrays that are aligned to the \acrfullpl{gc} and \acrfullpl{2dc}.
		The output state from the chip is analysed by two \acrfull{fst} units consisting of a \acrlong{qwp}, a \acrlong{hwp} and a \acrlong{pbs}, together with \acrfullpl{snspd}.
		\textbf{(b)} Zoom-in of the integrated photonic circuit.
		The integrated circuit is based on the \acrlong{soi} material platform, utilising a slot waveguide, with a width of $\SI{450}{nm}$ and a height of $\SI{220}{nm}$.
		On the left side of the circuit, four \glspl{gc} are placed to couple light into the four qubit modes (i10, i11, i20 and i21).
		Four \acrfullpl{mzi} consisting of, in total, six phase shifters (P11, P12, P21, P22, P23, PCR) and eight \acrlongpl{mmi} allow for active manipulation of both qubits.
		The output of the circuit is connected to two \glspl{2dc} (o1 and o2) on the right side.
		These \glspl{2dc} transform the path-encoded qubits to two polarisation-encoded qubits and emit them out of the chip plane.
		In each of the arms of the \glspl{2dc} a \gls{mzi} (P1H, P1V, P2H, P2V) is placed for active compensation of imbalances in the \glspl{2dc}.
		\textbf{(c)} Schematic of the Bell state source used in the on-chip tomography experiment.
		Here, the crystal is placed within a Sagnac-type interferometer and the pump beam is diagonally polarised to pump the crystal from both directions~\cite{Jin_2014}.
		In this configuration, the photons generated from both directions of pumping are indistinguishable, resulting in the generation of polarisation entangled photon pairs.
		The photons propagating along the pump beam are separated using a dichroic mirror and coupled into single-mode fibres.
		Additionally, $\SI{1.5}{nm}$ band-pass filters centered at $\SI{1550}{nm}$ are used to ensure spectral indistinguishability. 
		We connect each of the two photons of the source to one of the \glspl{2dc}.
		An \gls{snspd} channel is connected to each of the four \glspl{gc} i10, i11, i20 and i21.
		\textbf{(d)} Schematic of the setup for the chip-to-chip experiment.
		We place two copies of our chip in two individual coupling setups and connect them using single-mode fibres.
		Again, we use a linear \gls{spdc} source to generate a pair of single photons.
		Four \glspl{snspd} are utilised to detect a photon in each qubit mode.
		The left chip, connected to the photon source acts as a sending network node which prepares the entangled state and the right chip, connected to the \glspl{snspd} acts as the receiving network node which performs two-qubit \gls{fst}.
		Two in-fibre $\SI{1.5}{nm}$ band-pass filters centered at $\SI{1550}{nm}$ are placed in each fibre connecting the two chips.
	}
\end{figure*}
Our \gls{soi}-based photonic circuit features four waveguides acting as spatial modes.
To encode our qubits we use dual-rail encoding, thus our chip encodes two qubits.
Here, a photon in the upper waveguide corresponds to $\ket{0}$, and a photon in the lower waveguide to $\ket{1}$.
The integrated circuit features \acrfullpl{mzi} consisting of thermo-optic phase shifters and \acrfullpl{mmi}, allowing for active control of the chip to prepare arbitrary two-qubit states (see Fig.~\ref{fig:setup/fullsetup}).
A key element for converting path-encoded qubits to polarisation-encoded qubits, whilst conserving the qubit state, are \acrfullpl{2dc}, placed on the right side of the circuit~\cite{Rosa_2017}.
For this, light coming from the upper waveguide is emitted out of the chip plane horizontally polarised, and light coming from the lower waveguide is emitted vertically polarised, and vice versa when going from polarisation- to path-encoding, implementing the transformation
\begin{equation}
	\alpha\ket{0} + \beta\ket{1} \leftrightarrow \alpha\ket{H} + \beta\ket{V} \text{,}
\end{equation}
with $\alpha$ and $\beta$ being complex amplitudes (see Methods section for a more detailed analysis of the \glspl{2dc}' behaviour).

Pairs of single photons are generated in a \gls{spdc} source, based on \acrfull{ppktp} crystals (see Fig.~\ref{fig:setup/fullsetup} for details).
The single photons are coupled into fibres and into the photonic chip.
To analyse the state prepared on our chip, the output state is again collected by two fibers and analysed by two \gls{fst} units, which are connected to \acrfullpl{snspd}.
We exchange the linear photon source and the tomography stages for a Sagnac-type Bell state source to demonstrate the chip's capability to act as a two-qubit tomography unit.
In a third experiment, to fully demonstrate our chip's potential as a quantum network node, we connect two copies of our chip using single-mode fibres.
One chip, acting as a sending node, is connected to the linear \gls{spdc} source from our first experiment.
The second chip, acting as a receiving node, is connected to four \glspl{snspd}.
The photonic chip and the setup are presented in Fig.~\ref{fig:setup/fullsetup}.

\section{Results}

As a first test of our experimental setup, we perform a \gls{hom} experiment at the \gls{mzi} containing the phase shifter PCR by setting the splitting ratio to $\SI{50}{\%}$~\cite{Hong_1987}.
\begin{figure}
	\includegraphics[scale=0.48]{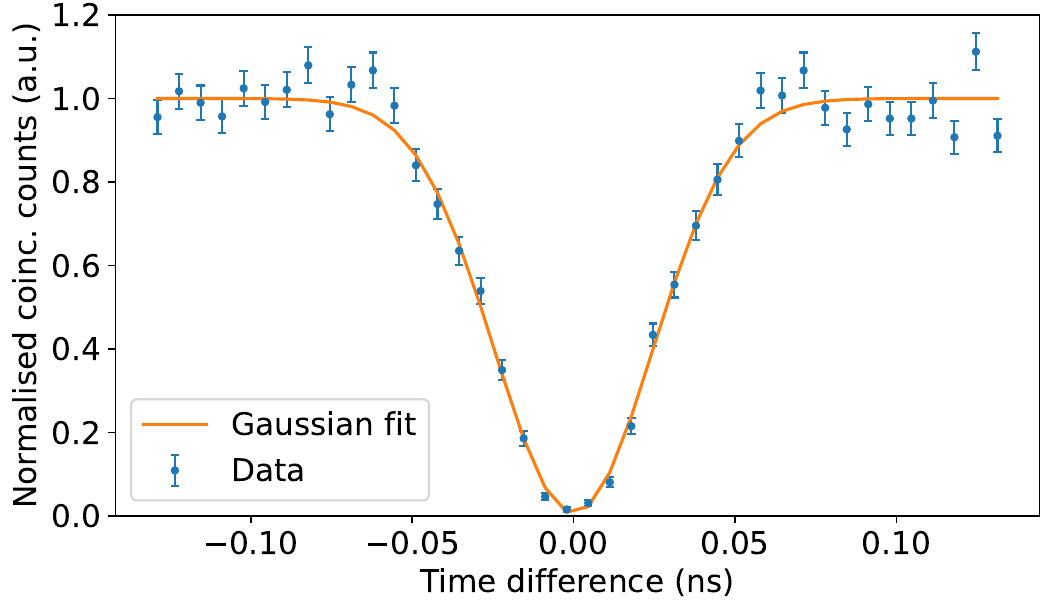}
	\caption{
        \label{fig:results/homdip}
        \Acrfull{hom} interference measured at the crossing \acrfull{mzi} (containing PCR) using the two photons emitted by the linear \acrlong{spdc} source shown in Fig.~\ref{fig:setup/fullsetup}.
		The temporal shift in arrival time is introduced by changing the path length to the MZI of one of the two photons by the off-chip delay line.
		We measure a \gls{hom} visibility of $\SI{99.5 +- 0.4}{\%}$.
	}
\end{figure}
From the measured data (see Fig.~\ref{fig:results/homdip}), we obtain a visibility of $\SI{99.5 +- 0.4}{\%}$, demonstrating the high quality of our photon source and the chip operation.

To demonstrate the functionality of the chip for state generation, we prepare three unentangled two-qubit states.
By setting $\phi_\text{PCR}$ and $\phi_\text{P23}$ accordingly (see Table~\ref{tab:results/phasesettings}), we completely separate the two on-chip qubits from each other, leaving an \gls{mzi} plus an outer phase shifter in each qubit.
When sending in a photon in the upper input, such an \gls{mzi} implements the transformation
\begin{equation}
    \ket{0}_k \rightarrow \alpha_k \ket{0}_k + \beta_k \ket{1}_k
\end{equation}
with $k\in\qty{1,2}$ labelling the qubit and $\alpha_k$ and $\beta_k$ being dependent on the phase settings as listed in Table~\ref{tab:results/phasesettings}.
We choose to prepare the eigenstates of the Pauli matrices $\ket{H,V}$, $\ket{+,-}$, and $\ket{+_\text{i},-_\text{i}}$ (see Table~\ref{tab:results/phasesettings} for detailed settings).
The states are then coupled out of the chip by the \glspl{2dc} and we perform \gls{fst} using the off-chip tomography stages (see Fig.~\ref{fig:setup/fullsetup}).
\begin{figure}
	\centering
	\includegraphics[scale=0.4]{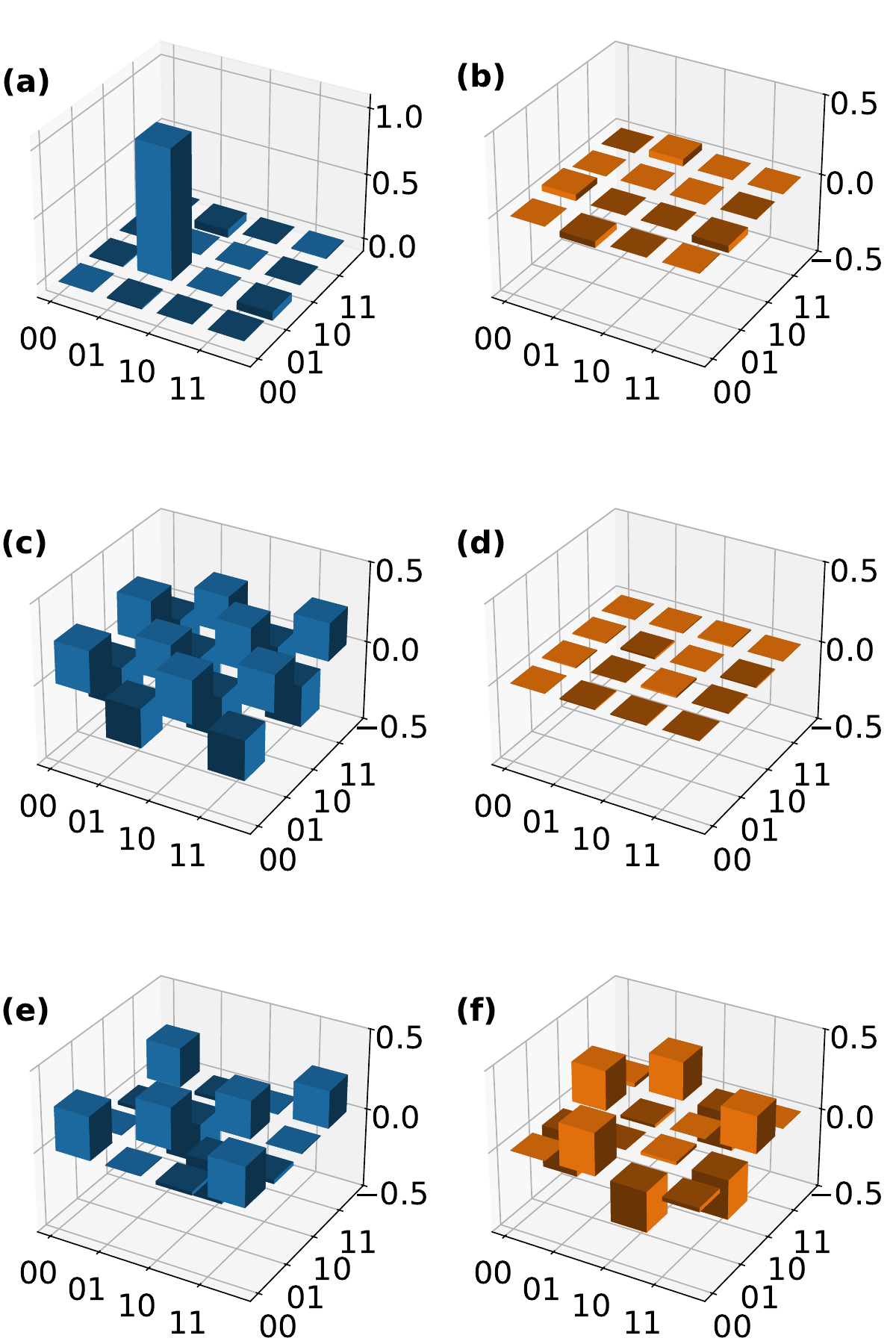}
	\caption{
        \label{fig:results/onchip_unentangledstates_rho}
        Real and imaginary parts of the reconstructed density matrices of \textbf{(a,b)} $\ket{H,V}$, \textbf{(c,d)} $\ket{+,-}$, and \textbf{(e,f)} $\ket{+_\text{i},-_\text{i}}$.
		The states are prepared on chip by setting the phases according to Table~\ref{tab:results/phasesettings} and coupled to a pair of optical fibres via the \acrlongpl{2dc}.
		Each photon is send to a \acrfull{fst} unit as shown in Fig.~\ref{fig:setup/fullsetup}.
		We perform \gls{fst} and recover the density matrix of each state by a \acrlong{mle} on the measured coincidence counts.
		The corresponding fidelities and purities of each state are listed in Table~\ref{tab:results/phasesettings}.
	}
\end{figure}
We perform \gls{mle} to analyse the results.
For the recovered density matrix $\rho_\text{exp}$, we calculate the fidelity $F$ using the equation
\begin{equation}
	F = \qty( \text{tr} \sqrt{\sqrt{\rho_\text{target}} \rho_\text{exp} \sqrt{\rho_\text{target}}} ) ^2
	\label{eq:fidelity}
\end{equation}
and the purity using
\begin{equation}
	P = \text{tr}\qty( \rho_\text{exp}^2 )
	\label{eq:purity}
\end{equation}
with $\rho_\text{target}$ being the theoretically expected density matrix.
This leads to an average fidelity of $\SI{99.319 +- 0.010}{\%}$ and an average purity of $\SI{99.41 +- 0.02}{\%}$.
The reconstructed density matrices of the three states are visualised in Fig.~\ref{fig:results/onchip_unentangledstates_rho}.

Following this, we prepare the set of all four Bell states on-chip.
Choosing the phase settings according to Table~\ref{tab:results/phasesettings} and post-selecting on one photon in each qubit mode, we prepare
\begin{equation}
	\begin{aligned}
    \ket{00} &\rightarrow \qty(1+\alpha)\qty(\ket{00}-\beta\ket{11}) \,\text{.}\\
	&+ \qty(1-\alpha)\qty(\ket{01}+\beta\ket{10}) 
	\end{aligned}
\end{equation}
Again, we couple the on-chip prepared states out of the chip using the \glspl{2dc} and perform off-chip \gls{fst}.

\begin{figure*}
	\includegraphics[scale=0.4]{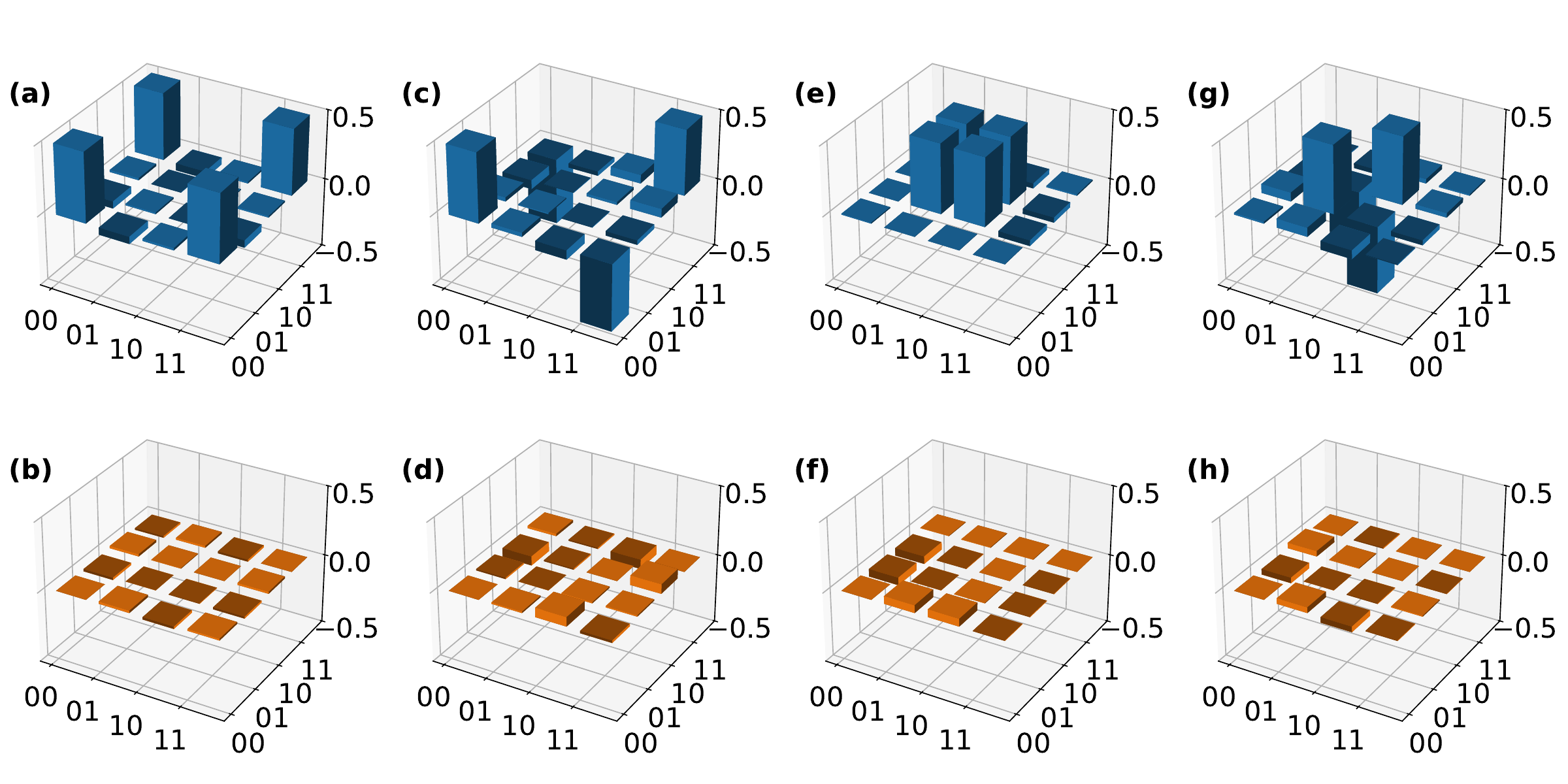}
	\caption{
        \label{fig:results/onchip_bellstates_rho}
        Real and imaginary part of the recovered density matrices of \textbf{(a,b)} $\ket{\Phi^\text{+}}$, \textbf{(c,d)} $\ket{\Phi^\text{-}}$, \textbf{(e,f)} $\ket{\Psi^\text{+}}$, and \textbf{(g,h)} $\ket{\Psi^\text{-}}$.
		The state is prepared on the chip by setting the phases according to Table~\ref{tab:results/phasesettings} and coupled to a pair of optical fibres via the \acrlongpl{2dc}.
		Each photon is send to a \acrfull{fst} unit as shown in Fig.~\ref{fig:setup/fullsetup}.
		We perform \gls{fst} and recover the density matrix of each state by a \acrlong{mle} on the measured coincidence counts.
		The corresponding fidelities and purities of each state are listed in Table~\ref{tab:results/phasesettings}.
	}
\end{figure*}
The measured states' density matrices are presented in Fig.~\ref{fig:results/onchip_bellstates_rho} and we determine an average fidelity of $\SI{97.69 +- 0.03}{\%}$, and an average purity of $\SI{97.93 +- 0.07}{\%}$.
The error is estimated from Poissonian distributed coincidence counts.
These fidelities are, to our knowledge, the highest reported fidelities of an on-chip prepared Bell state measured off-chip.

\begin{table*}
    \caption{
        \label{tab:results/phasesettings}
        Phase settings for the on-chip preparation of all six eigenstates of the three Pauli operators and all four Bell states.
		The phase shifters P1H, P1V, P2H, and P2V are set to compensate for imbalances introduced by the \acrfullpl{2dc} during the experiments.
		The chip layout and phase shifter locations are shown in Fig.~\ref{fig:setup/fullsetup}.
		The states prepared on-chip are coupled out of the chip using \glspl{2dc} and analysed by two off-chip \acrlong{fst} units.
		From the measured coincidence counts, the density matrix of each state is recovered using \acrlong{mle}.
		Here, the determined fidelity (see Eq.~\ref{eq:fidelity}) and purity (see Eq.~\ref{eq:purity}) of each state are given.
		Additionally, the phase settings to prepare a two-qubit cluster state $\ket{C}$, which is sent from the sender chip to the receiver chip, and the determined fidelity and purity are given.
		We also list the phase settings for setting the on-chip bases $X$, $Y$, and $Z$.
    }
	\begin{tabular*}{\textwidth}{@{\extracolsep{\fill}}lcccccccc}
		\toprule
        State/Basis & $\phi_\text{P11}$ & $\phi_\text{P12}$ & $\phi_\text{P21}$ & $\phi_\text{P22}$ & $\phi_\text{PCR}$ & $\phi_\text{P23}$ & Fidelity $\qty(\unit{\%})$ & Purity $\qty(\unit{\%})$\\
		\midrule
		$\ket{H,V}$ & $\pi$ & $0$ & $0$ & $0$ & $\pi$ & $\pi$ & $\num{99.274 +- 0.015}$ & $\num{99.98 +- 0.03}$ \\
		$\ket{+,-}$ & $\pi/2$ & $\pi$ & $\pi/2$ & $0$ & $\pi$ & $\pi$ & $\num{99.553 +- 0.014}$ & $\num{99.40 +- 0.03}$ \\
		$\ket{+_\text{i},-_\text{i}}$ & $\pi/2$ & $3\pi/2$ & $\pi/2$ & $\pi/2$ & $\pi$ & $\pi$ & $\num{99.13 +- 0.02}$ & $\num{98.84 +- 0.05}$ \\
        $\ket{\Phi^+}$ & $\pi/2$ & $\pi$ & $\pi/2$ & $0$ & $0$ & $0$ & $\num{98.13 +- 0.07}$ & $\num{97.70 +- 0.14}$ \\
        $\ket{\Phi^-}$ & $\pi/2$ & $0$ & $\pi/2$ & $0$ & $0$ & $0$ & $\num{97.00 +- 0.07}$ & $\num{98.09 +- 0.15}$ \\
        $\ket{\Psi^+}$ & $\pi/2$ & $0$ & $\pi/2$ & $0$ & $0$ & $\pi$ & $\num{98.25 +- 0.06}$ & $\num{98.36 +- 0.13}$ \\
        $\ket{\Psi^-}$ & $\pi/2$ & $\pi$ & $\pi/2$ & $0$ & $0$ & $\pi$ & $\num{97.39 +- 0.07}$ & $\num{97.57 +- 0.15}$ \\
		$\ket{C}$ & $\pi/2$ & $0$ & $\pi/2$ & $0$ & $0$ & $\pi/2$ & $\num{90.0 +- 1.6}$ & $\num{91 +- 3}$ \\
		$XX$ & $\pi/2$ & $0$ & $\pi/2$ & $0$ & $\pi$ & $\pi$ & - & - \\
		$YY$ & $\pi/2$ & $\pi/2$ & $\pi/2$ & $\pi/2$ & $\pi$ & $\pi$ & - & - \\
		$ZZ$ & $\pi$ & $0$ & $\pi$ & $0$ & $\pi$ & $\pi$ & - & - \\
		\bottomrule
	\end{tabular*}	
\end{table*}

\begin{figure}
	\includegraphics[scale=0.4]{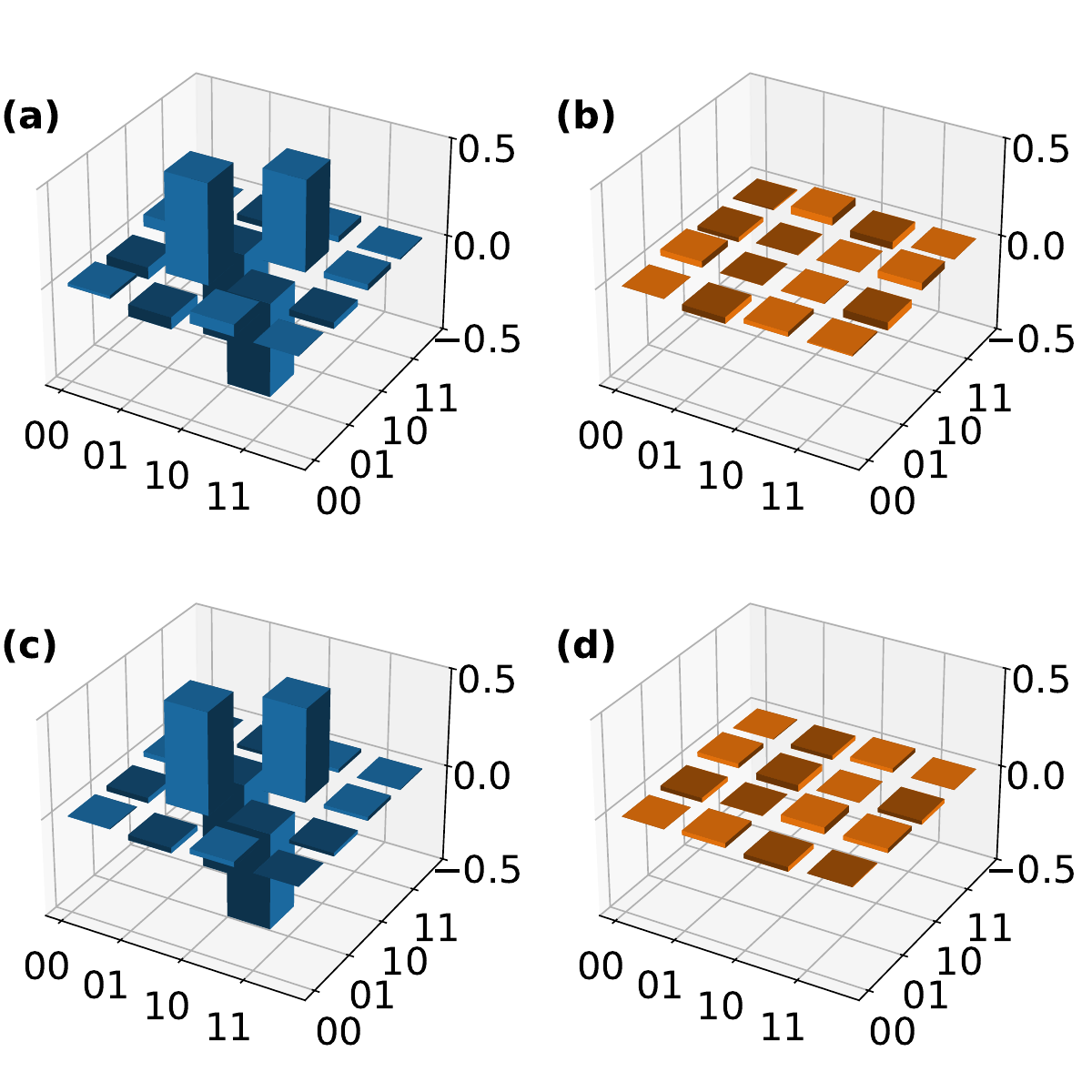}
	\caption{
        \label{fig:results/onchip_tomo_rho}
        We prepare the state $\ket{\Psi^-}$ off-chip and couple it into the photonic chip, with the chip functioning as two \acrfull{fst} units.
		We perform \gls{fst} on the state and analyse the results by \acrlong{mle}.
		Here, the \textbf{(a)} real and \textbf{(b)} imaginary parts of the recovered density matrix are visualsed.
		The fidelity of the measured state is $\SI{97.3 +- 0.2}{\%}$ and its purity is $\SI{97.6 +- 0.5}{\%}$.
		As a benchmark for the on-chip \gls{fst} units, we also perform off-chip \gls{fst} on the prepared Bell state.
		The recovered density matrix' \textbf{(c)} real and \textbf{(d)} imaginary parts are visualised.
		We determine a fidelity of $\SI{99.109 +- 0.006}{\%}$ and a purity of $\SI{99.321 +- 0.015}{\%}$.
	}
\end{figure}
The chip's functionality is not limited to preparing any arbitrary one- or two-qubit state, but it is also capable of measuring any one- or two-qubit state.
To demonstrate this, we utilise the chip as a two-qubit \gls{fst} unit for two-qubit states that are prepared off-chip in polarisation-encoding and then sent to the chip, where the encoding is changed to path-encoding.
For that, we utilise an \gls{spdc} source in a Sagnac configuration, which generates the state $\ket{\Psi^-}$ (see Fig.~\ref{fig:setup/fullsetup}).
By adjusting the phase shifters according to Table~\ref{tab:results/phasesettings}, we can project each qubit onto one of the three bases $X$-, $Y$-, and $Z$.
The density matrix of the measured state, is visualised in Fig.~\ref{fig:results/onchip_tomo_rho} and exhibits a fidelity of $\SI{97.3 +- 0.2}{\%}$ and a purity of $\SI{97.6 +- 0.5}{\%}$.
We benchmark our result by an off-chip \gls{fst} of the same state, yielding a fidelity of $\SI{99.109 +- 0.006}{\%}$ and a purity of $\SI{99.321 +- 0.015}{\%}$.
This is in good agreement with the on-chip measurement with a relative fidelity
\begin{equation}
	F_\text{rel} = \frac{F_\text{on-chip}}{F_\text{off-chip}} = \SI{98.2 +- 0.2}{\%} \text{,}
\end{equation}
demonstrating the functionality of our chip as a receiving unit in a quantum network.
The difference in fidelity between the measurements is likely caused by non-unitary cross-talk in the \gls{2dc}, which reduces the fidelity of the state in an irreversible way.

\begin{figure}
	\includegraphics[scale=0.4]{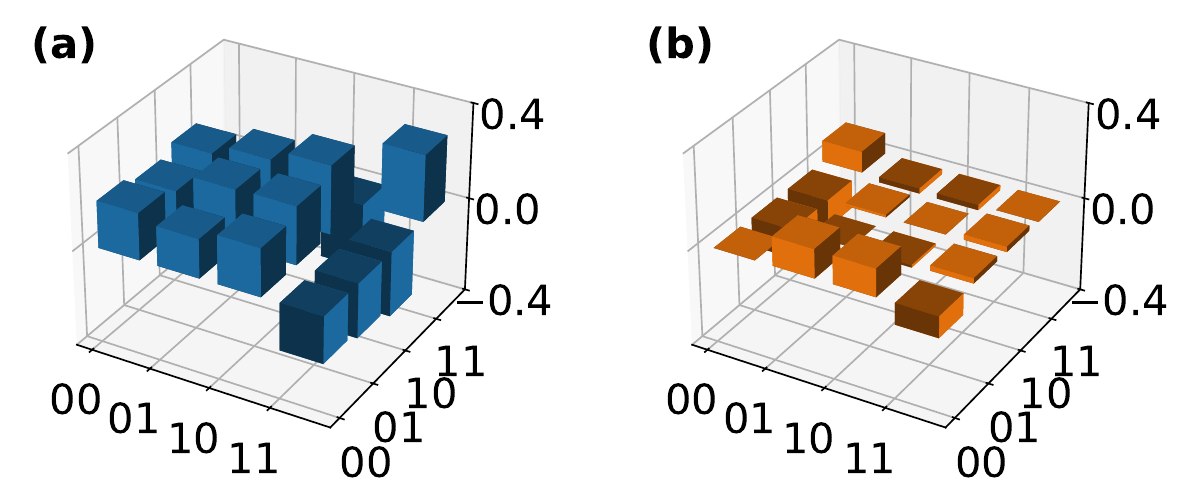}
	\caption{
        \label{fig:results/chiptochip_rho}
        Using the sender chip we prepare a two-qubit cluster state $\ket{C}$ and send it to the receiver chip.
		Using the receiver chip, we perform on-chip \acrlong{fst} on the state and analyse the results by \acrlong{mle}.
		Here, the \textbf{(a)} real and \textbf{(b)} imaginary parts of the recovered density matrix are visualsed.
		The fidelity of the measured state is $\SI{90.0 +- 1.6}{\%}$ and its purity is $\SI{91 +- 3}{\%}$.
	}
\end{figure}
Having shown the potential of our circuit to act as a sending and as a receiving unit, we demonstrate its application as a flexible quantum network node by connecting two copies of our chip (see Fig.~\ref{fig:setup/fullsetup}).
The first chip acts as a sending unit by preparing the two-qubit cluster state
\begin{equation}
	\ket{C} = \frac{1}{4} \qty( \ket{00} + \ket{10} + \ket{01} - \ket{11} )
\end{equation}
on chip.
The output photons are connected to a second copy of the chip, which acts as a receiving unit and performs measurements in combinations of $X$, $Y$ and $Z$ bases.
We perform on-chip \gls{fst} using the receiver chip and use \gls{mle} to analyse the results.
All phase settings are given in Table~\ref{tab:results/phasesettings}.
The recovered density matrix is visualised in Figure~\ref{fig:results/chiptochip_rho}.
We determine a fidelity of $\SI{90.0 +- 1.6}{\%}$ and a purity of $\SI{91 +- 3}{\%}$, which confirms the successful transmission of a maximally entangled state from one chip to another.

\section{Conclusion}
In this work, we present a photonic chip that is capable of preparing arbitrary one- or two-qubit states and can act as a measurement unit for up to two qubits.
We demonstrate its state preparation functionality by preparing three different unentangled two-qubit states and the set of all four Bell states.
For the unentangled states, we measure an average fidelity of $\SI{99.319 +- 0.010}{\%}$, while the average fidelity for the Bell states is measured to be $\SI{97.69 +- 0.03}{\%}$.
Furthermore, we use the chip to perform \gls{fst} on an off-chip prepared Bell state and measure a fidelity of $\SI{97.3 +- 0.2}{\%}$.
We combine these two operational modes of our chip by connecting two copies of it, realising a proof-of-concept quantum network.
We prepare a maximally entangled state on the first chip and perform \gls{fst} on the second, achieving a fidelity of $\SI{90.0 +- 1.6}{\%}$ and therefore proving that our chip is capable of sharing entanglement across the network.

The demonstrated combined functionality of state preparation and detection with high fidelities makes our chip a versatile node in a quantum network.
Due to our chip being low in power consumption and able to prepare Bell states, it is also an ideal candidate for entanglement-based quantum key distribution protocols, for example, as a sender unit on a satellite.
In addition, our approach allows for scaling to a higher number of qubits by adapting the state preparation circuit.
Future steps include the cointegration of on-chip photon sources and detectors with our chip, enabling a fully integrated operation of the quantum network node.

\bibliography{bibfile.bib}

\section{Acknowledgements}
We thank Christian Schweikert and Wolfgang Vogel for providing the design of the two-dimensional grating couplers.
We also thank Simone E. D'Aurelio for support with the measurement software.
We acknowledge support from the Carl Zeiss Foundation, the Centre for Integrated Quantum Science and Technology (IQST), the Federal Ministry of Research, Technology and Space (BMFTR, projects SiSiQ: FKZ 13N14920, PhotonQ: FKZ 13N15758, QRN: FKZ16KIS2207), and the Deutsche Forschungsgemeinschaft (DFG, German Research Foundation, 431314977/GRK2642, SFB 1667).

\section{Disclosures}
The authors declare no competing interests.

\section{Methods}

\subsection{Imperfections of 2DGCs}
\label{sec:methods/2dc}
\begin{figure}[H]
	\includegraphics[scale=0.34]{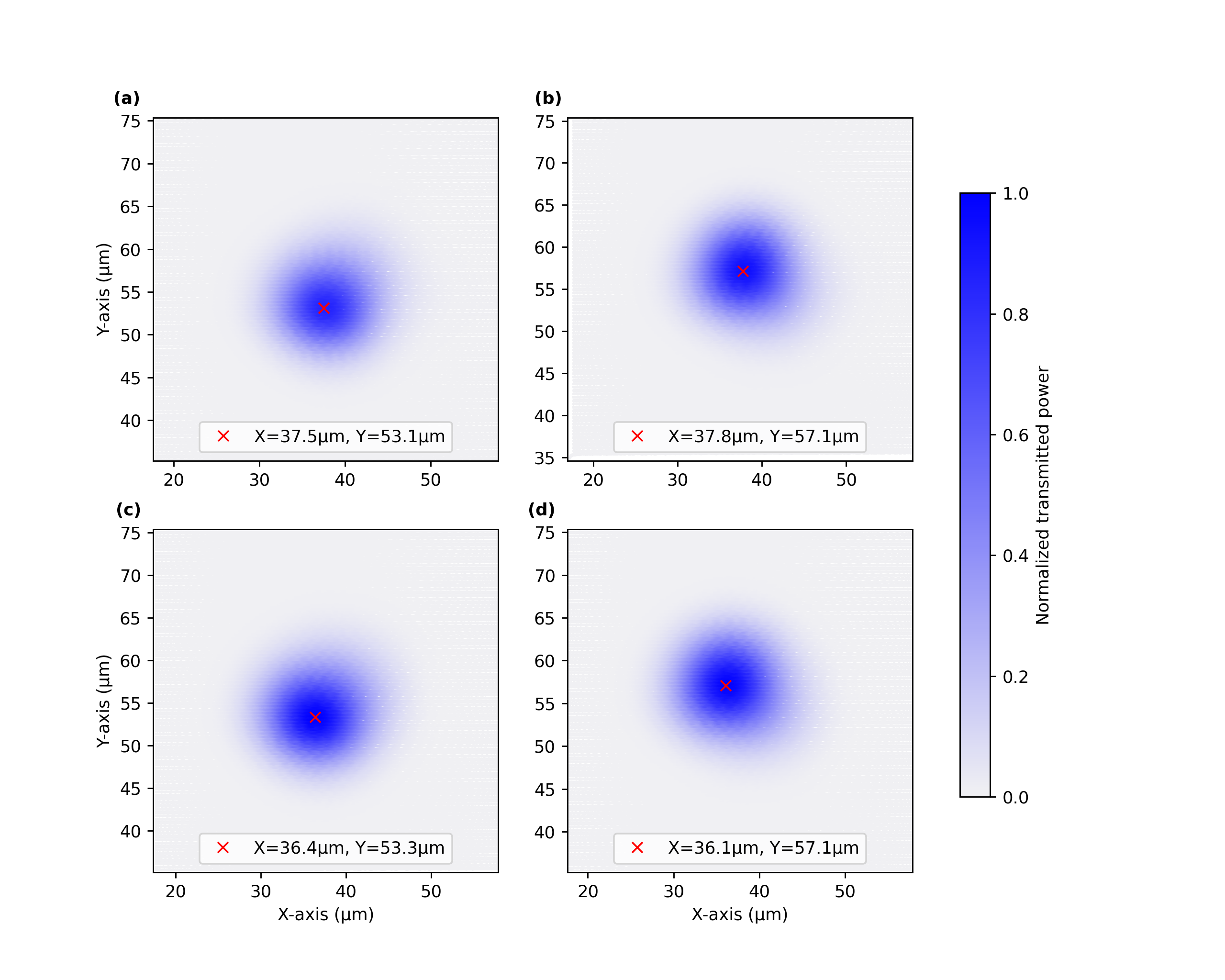}
	\caption{
        \label{fig:appendix/2dgc/areascans}
        Area scans of the emitted spatial light mode of both \acrfullpl{2dc}.
		A classical $\SI{1550}{nm}$ light source is coupled into the \textbf{(a)} upper and \textbf{(b)} lower arm of the first \gls{2dc} and into the \textbf{(c)} upper and \textbf{(d)} lower arm of the second \gls{2dc}.
		The red cross indicates the intensity maximum.
	}
\end{figure}

The two \glspl{2dc} transform the photon's state from the path-encoding to the polarisation-encoding, they thus implement the transformation
\begin{equation}
\begin{aligned}
    \ket{0}&\leftrightarrow\ket{H}\\
	\ket{1}&\leftrightarrow\ket{V}
\end{aligned}
\end{equation}
with $\ket{0}$ and $\ket{1}$ referring to path-encoded states and $\ket{H}$ and $\ket{V}$ to polarisation-encoded states.
Furthermore, for an ideal \gls{2dc}, we expect both polarisations to be emitted from the coupler into the same spatial position, where we place a fibre to couple in the photons.
Our experimental characterisation shows a small deviation from the ideal behaviour. 
From our data, we estimate a cross-talk of about $\SI{1}{\%}$, corresponding to a path-encoded state being converted into the wrong polarisation.
In addition, the two spatial modes corresponding to the two polarisation states slighlty deviate from eachother (see Fig.~\ref{fig:appendix/2dgc/areascans}). 
We compensate for this imbalance in coupling efficiency by introducing additional \glspl{mzi}.

\end{document}